\documentclass[12pt]{article}
\usepackage[centertags]{amsmath}

\begin{document}
\title{Second comment to ``Invariance of the tunneling method''}
\date{\today}
\author{Marco Pizzi\\Physics Department, University of Rome ``La Sapienza'', \\P.le A. Moro, Italy 00185, \\E-mail: pizzi@icra.it}

\maketitle
\begin{abstract}
I reply to the four points raised by S. A. Hayward, R. Di Criscienzo, M. Nadalini, L. Vanzo, S. Zerbini (arXiv:0909.2956v1) against my comment (arXiv:0907.2020v1) to their previous article. I maintain my position on the wrongness of their paper, reporting also another mistake.
\end{abstract}

\section{Reply to Ref.\cite{DHNVZ2}}
   Nowhere in my papers I said that the Hawking radiation itself ``is a myth" or something equivalent. I simply said that it can not be explained by the way proposed in the articles \cite{DHNVZ,NVZ,Ang} (and in a number of analogous papers) \emph{because these articles are wrong}. I call the reader to not mix this very concrete statement and the general problem regarding the existence of the Hawking radiation. Even if it exists there is no reason to accept any false manipulation pretending to be a ``derivation", and at least my argument can be considered as ``the confutation of a wrong demonstration".  
  
   The authors of \cite{DHNVZ2} mentioned four points against my comment \cite{Pizzi2}; for the ease of the reader I report here their words:
   
\begin{enumerate}
\item  The author states that ``the action...along the classical light-like ray is...constant" and therefore ``no imaginary part in the action can appear". However, Hawking radiation is not a prediction of classical physics but of quantum field theory. In the WKB approximation, the action is not constant, but indeed rapidly varying.
\item The author states that an ``infinitesimally small neighbourhood of the horizon...can be covered by Minkowski coordinates". This is incorrect. The correct statement is that connection coefficients, being first derivatives of the metric, can be set to zero at a point. Surface gravity and the corresponding temperature are
curvature invariants, which involve second derivatives of the metric and cannot be set to zero at a point.
\item There is a confusion of partial derivatives in the author's equation (3). He appears to be solving the null geodesic equation rather than the Hamilton-Jacobi equation.
\item The author's procedure for dealing with the pole in the action is inequivalent to the standard one, namely the Feynman $i\epsilon$ procedure or something equivalent, which
corresponds to the desired physical boundary conditions. It has not been justified and is used by no other author as far as we are aware.
\end{enumerate}

My reply is:
\begin{enumerate}
\item Of course, ``Hawking radiation is not a prediction of classical physics but of quantum field theory". However, even in quantum theory in the principal WKB approximation the main contribution to the action is given by the classical (although complexified) trajectory, and along this trajectory the action of any massless particle is a constant. To affirm that because of quantum theory we will have in the principal WKB approximation an action of essentially different character is non-sense.

\item  Nobody objects the assertion that an invariant can not be canceled by a coordinate choice. Nevertheless this trivial fact has nothing in common with my statement. My assertion is that the calculation of the action in papers \cite{DHNVZ,NVZ,Ang} is erroneous and if calculated correctly it does not contains any invariant term depending on the second derivatives of the metric (see my \cite{Pizzi2,Pizzi}). This is evident because the alteration of the action along an infinitesimally small segment of a geodesics is (up to a constant factor) the relativistic invariant interval $ds$ (or the differential of an appropriate parameter in case of a massless particle). This quantity is defined only by the equation of geodesics (i.e. only by the metric and connection coefficients) and its limit for an infinitesimal interval of the segment \emph{in no way depends on the second derivatives of the metric}. Furthermore, since $ds$ evidently does not contains any imaginary part in the locally inertial system nearby a regular space-time point (as any point of horizon is), then it can not contains such part also in any other coordinate system (being $ds$ invariant), no matter of any other curvature invariant. If one obtained a result which depends on these second derivatives this means simply that the action was not correctly estimated; in the next section I briefly show another of these mistakes present in \cite{DHNVZ}.

\item To find the equation for the trajectory one can use either geodesic equation or Hamilton-Jacobi equation. There is no difference. The equation (3) in my \cite{Pizzi2} is a trivial identity which in no way can contain any confusion, anybody can check it in one minute.

\item  There are \emph{no poles} in the integrand of the action, and this is the main conclusion of my papers. Then any discussion on how to deal with the pole is irrelevant.
\end{enumerate}

Finally, although my arguments hold for any black hole with a regular horizon (dynamical or not), in order to judge on the correctness of my statements I suggest the reader to focus firstly on the simpler case of the Schwarzschild black hole, where only very-well-known formulas are used and any eventual mistake is easier to be discovered.

\section{Another mistake}
Analyzing the paper \cite{DHNVZ} is possible to find also another remarkable mistake. Indeed in Section 5.4, trying to recover the same result of their previous sections (now using the ``Lema\^{\i}tre-Rylov gauge'' for the Schwarzschild black hole), they did a simple (but fatal) algebraic error. From their formula (V.65)\footnote{Also this formula (V.65) is wrong since $\omega$ should be replaced by $\omega/\sqrt{B}$, as one can see from (V.64); but this is a blind error for our purpose because $B=1$ on the horizon.} it is immediate to see that the consequent (V.68) has the wrong sign in front of $\omega$. After correcting this mistake, it is easy to show that the sum of the temporal part of the action and the radial one gives not their expression (V.72) but $\texttt{Im}[I_+]$ identically equal to zero. Therefore in this case they involuntarily confirmed my previous result.

\section*{Final remark}
I just wish to underline once more that I confined myself to the semiclassical approximation, and \emph{in this framework} there is no satisfactory explanation, up to now, of the original Hawking's result. Thus, in my opinion, the question is open and the problems raised in \cite{Bel06}-\cite{Bel07} need further investigation.

\end{document}